\documentclass[aps,prl,reprint,graphicx,amsmath, superscriptaddress, showpacs]{revtex4-1}
\usepackage{graphicx}
\usepackage{gensymb}

\draft 

\begin{document}


\title{High-brilliance betatron gamma-ray source powered by laser-accelerated electrons} 

\author{J. Ferri}
\email{julien.ferri@polytechnique.edu}
\affiliation{CEA, DAM, DIF, 91297 Arpajon, France}
\affiliation{LOA, ENSTA ParisTech, CNRS, Ecole Polytechnique, Universit\'e Paris-Saclay, 91762 Palaiseau, France}
\author{S. Corde}
\affiliation{LOA, ENSTA ParisTech, CNRS, Ecole Polytechnique, Universit\'e Paris-Saclay, 91762 Palaiseau, France}
\author{A. D\"opp}
\affiliation{LOA, ENSTA ParisTech, CNRS, Ecole Polytechnique, Universit\'e Paris-Saclay, 91762 Palaiseau, France}
\affiliation{Ludwig-Maximilians-Universit\"at M\"unchen, Fakult\"at f\"ur Physik, Am Coulombwall 1, Garching 85748, Germany}
\author{A. Lifschitz}
\affiliation{LOA, ENSTA ParisTech, CNRS, Ecole Polytechnique, Universit\'e Paris-Saclay, 91762 Palaiseau, France}
\author{A. Doche}
\affiliation{LOA, ENSTA ParisTech, CNRS, Ecole Polytechnique, Universit\'e Paris-Saclay, 91762 Palaiseau, France}
\author{C. Thaury}
\affiliation{LOA, ENSTA ParisTech, CNRS, Ecole Polytechnique, Universit\'e Paris-Saclay, 91762 Palaiseau, France}
\author{K. Ta Phuoc}
\affiliation{LOA, ENSTA ParisTech, CNRS, Ecole Polytechnique, Universit\'e Paris-Saclay, 91762 Palaiseau, France}
\author{B. Mahieu}
\affiliation{LOA, ENSTA ParisTech, CNRS, Ecole Polytechnique, Universit\'e Paris-Saclay, 91762 Palaiseau, France}
\author{I. Andriyash}
\affiliation{Synchrotron SOLEIL, L'Orme des Merisiers, Saint Aubin, 91192 Gif-sur-Yvette, France}
\affiliation{Department of Physics and Complex Systems, Weizmann Institute of Science, Rehovot, Israel}
\author{V. Malka}
\affiliation{LOA, ENSTA ParisTech, CNRS, Ecole Polytechnique, Universit\'e Paris-Saclay, 91762 Palaiseau, France}
\affiliation{Department of Physics and Complex Systems, Weizmann Institute of Science, Rehovot, Israel}
\author{X. Davoine}
\affiliation{CEA, DAM, DIF, 91297 Arpajon, France}

\date{\today}

\begin{abstract}
Recent progress in laser-driven plasma acceleration now enables the acceleration of electrons to several gigaelectronvolts. Taking advantage of these novel accelerators, ultra-short, compact and spatially coherent X-ray sources called betatron radiation have been developed and applied to high-resolution imaging. However, the scope of the betatron sources is limited by a low energy efficiency and a photon energy in the 10's of kiloelectronvolt range, which for example prohibits the use of these sources for probing dense matter. Here, based on three-dimensional particle-in-cell simulations, we propose an original hybrid scheme that combines a low-density laser-driven plasma accelerator with a high-density beam-driven plasma radiator, and thereby considerably increases the photon energy and the radiated energy of the betatron source. The energy efficiency is also greatly improved, with about 1\% of the laser energy transferred to the radiation, and the gamma-ray photon energy exceeds the megaelectronvolt range when using a 15~J laser pulse. This high-brilliance hybrid betatron source opens the way to a wide range of applications requiring MeV photons, such as the production of medical isotopes with photo-nuclear reactions, radiography of dense objects in the defense or industrial domains and imaging in nuclear physics.
\end{abstract}


\maketitle 

In laser wakefield acceleration (LWFA), large accelerating fields -- above 100~GV/m -- can be produced in the wake of an ultra-short and intense laser pulse as it propagates in an under-dense plasma, and can lead to the production of high-energy electron beams in very short distances~\cite{taji79, malk02}. The most efficient way to accelerate the electron beam is in the blowout -- also called bubble  -- regime~\cite{pukh02, lu06}. In this regime, the first period of the plasma wave driven by the laser pulse takes the form of an ion cavity surrounded by plasma electrons expelled by the ponderomotive force of the laser pulse. The accelerating and focusing fields in the ion cavity are ideal for the acceleration of electrons, and electron beams are now routinely accelerated to multi-GeV energies in cm-scale plasmas~\cite{wang13,leem14}. Besides, during their acceleration, electrons wiggle transversely and naturally emit synchrotron-like X-rays, known as betatron radiation~\cite{esar02, rous04}. This source has a broadband spectrum, that quickly drops after the critical photon energy $E_c=\hbar\omega_c\propto \gamma^2n_er_{\beta}$, where $n_e$ is the plasma density, $\gamma$ is the Lorentz factor of the electrons and $r_{\beta}$ is the amplitude of their transverse motion. Critical energies of tens of keV have previously been reported from betatron sources using laser energies of a few Joules to tens of Joules~\cite{wang13, knei10}. In addition, betatron radiation benefits from a micrometric size and a femtosecond duration, which makes it very interesting for applications requiring high-resolution diagnosis~\cite{cole15, wenz15, albe14}. Being perfectly synchronized, such femtosecond X-ray flashes are extremely well adapted for pump-probe experiments such as ultra-fast absorption spectroscopy.

\begin{figure*}[!t!]
\centering
\includegraphics[width=0.9\textwidth]{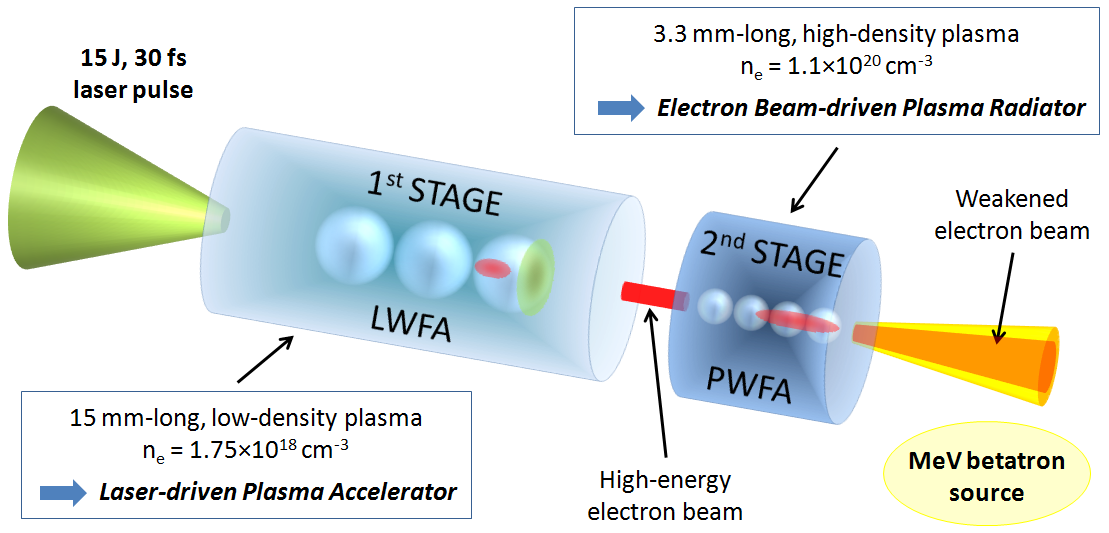}
\caption{Two-stage hybrid scheme for the production of a MeV betatron source. A 15~J, 30~fs (FWHM) laser pulse is focused at the entrance of a low-density gas cell. There, it excites a wakefield in the blowout regime and generates a high-energy electron bunch (laser wakefield regime, LWFA). After this first cell, the laser pulse is depleted and the electron bunch is sent on a second gas cell at a much higher density. The electron bunch drives a strong wakefield and experience strong transverse oscillations (plasma wakefield regime, PWFA), leading to the emission of an energetic photon beam in the $\gamma$-ray domain.}
\label{scheme}
\end{figure*}

However, the photon energy range accessible with these sources is limited to a few tens of keV, restraining its applications. Additionally, the energy transfer efficiency from the laser to the emitted radiation is so far of the order of $10^{-6}$ and is still to be improved in the experiments. The optimization of betatron sources in a laser wakefield accelerator indeed faces a major issue. On the one hand, in the blowout regime, the wakefield excited by the laser pulse propagates at approximately the laser group velocity $v_g = c\sqrt{1-\omega_p^2/\omega_0^2}$, close but substantially smaller than the speed of light $c$, where $\omega_0$ and $\omega_p=\sqrt{n_e e^2/\varepsilon_0 m_e}$ are respectively the laser and plasma frequencies, $e$ is the electron charge, $\varepsilon_0$ the vacuum permittivity and $m_e$ the electron mass. The electron beam in the wakefield has a velocity very close to $c$ and, thus, dephases with respect to the wakefield: it quickly overtakes the center of the ion cavity where it starts to experience a decelerating electric field. This occurs after a propagation distance called the dephasing length $L_\mathrm{deph} = (2\omega_0^2/3\omega_p^2)w_0$~\cite{lu07}, where $w_0$ is the laser spot size. For this reason, accelerating electrons to high energies~\cite{mart10,schr10} requires laser propagation at low plasma densities, where $L_\mathrm{deph}$ is higher. On the other hand, betatron emission is enhanced by a strong transverse wiggling and a short oscillation period, which preferentially happens in high-density plasmas. Consequently, the density $n_e$ can not be chosen to simultaneously optimize a high energy gain and a strong wiggling in a single stage, which severely limits the performance of the betatron source. To overcome this challenge, we propose in this Letter an original two-stage hybrid scheme, in which the acceleration process and the betatron emission are decoupled in two successive steps (see Fig.~\ref{scheme}). In a first stage, the laser pulse is sent in a low-density plasma, where the electron acceleration can be fully optimized. The generated electron beam is then sent to a second stage, which has a much higher plasma density. There, betatron radiation is emitted in a plasma wakefield accelerator~\cite{wang02} (PWFA), where the electron beam is driving the plasma wake~\cite{chen85, rose91}. No additional source of energy is required, as the plasma wake is powered by the electron beam. Moreover, a high density can be chosen because there is no problem of dephasing in this regime. This considerably enhances the emission in the beam-driven stage and the spectral range is then extended to the MeV level. Our results show that a 140~mJ photon beam with a critical energy of 9~MeV can be obtained from a 500~TW laser pulse, together with a very high brilliance $B>4\times 10^{23}~$photons/s/mm$^2$/mrad$^2$/0.1\%BW.

\begin{figure*}[!t!]
\centering
\includegraphics[width=0.9\textwidth]{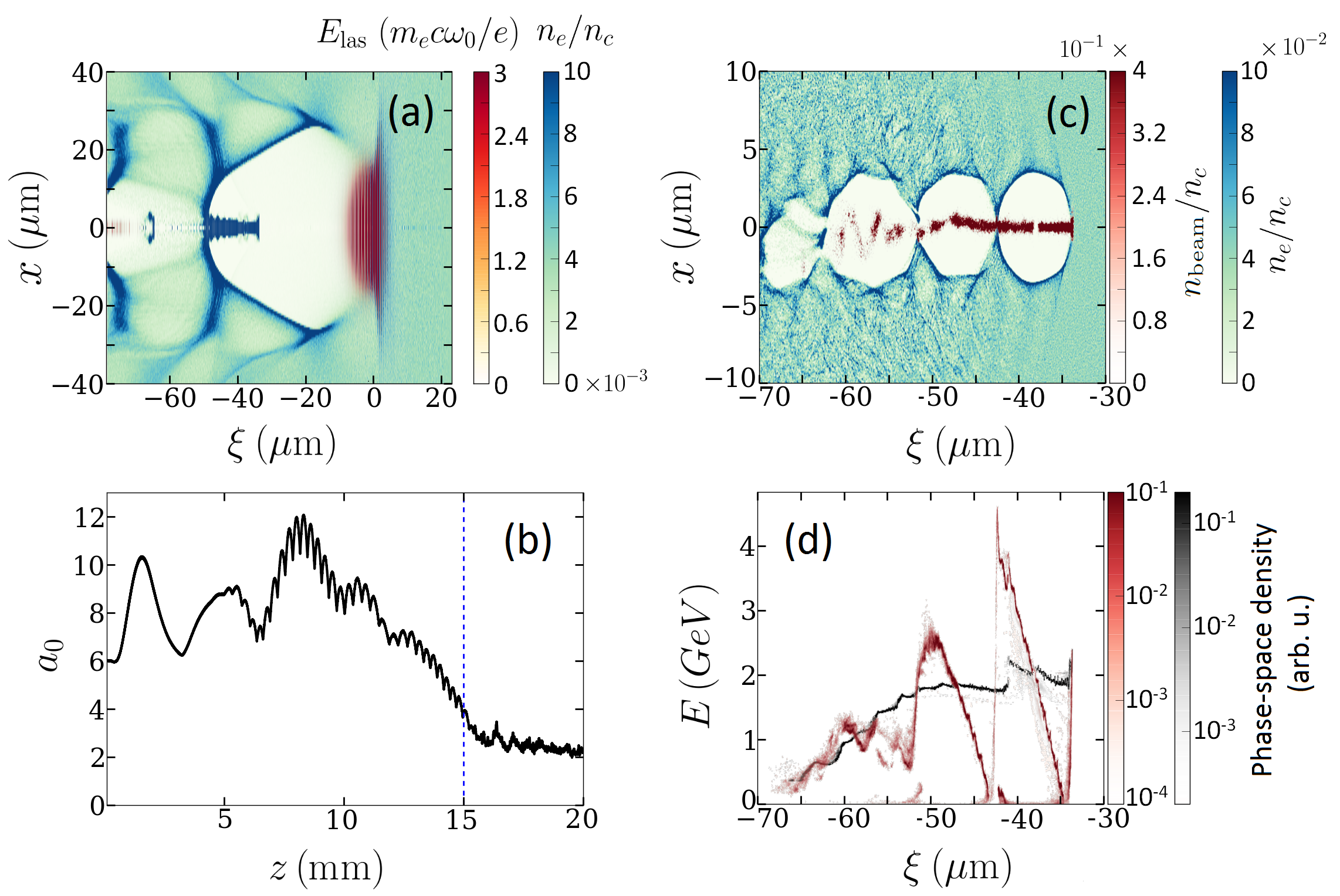}
\caption{Acceleration of electrons in the two-stage scheme. (a-b) Propagation in the LWFA regime (CALDER-Circ simulation): maps of the plasma density $\rho (|x|,z) = \rho(r,z)$ (green-blue) and of the laser field $E_\mathrm{las}(|x|,z) = E_\mathrm{las}(r,z)$ (yellow-red) after 5~mm of propagation (a) and evolution of the normalized peak potential vector $a_0$ of the laser with the distance of propagation (b). The electron beam is extracted from the CALDER-Circ simulation at the distance indicated with a dashed blue line in (b). (c-d) Propagation in the PWFA regime (3D CALDER simulation): maps of the plasma density (green-blue) and of the electron beam density (red) after 140~$\mu$m of propagation  (c) and longitudinal phase-space of the electron beam (d), at the beginning of the 3D simulation (gray) and after $740~\mu$m of propagation (red). In (a) and (c), densities are normalized to the critical density $n_c = \omega_0^2 m_e\varepsilon_0/e^2$, $x$ is the dimension transverse to the motion of the laser and the electron beam, and $\xi = z-ct$ is the dimension parallel to the motion, comoving at the speed of light in the direction of motion. In both simulations, $\xi = 0$ indicates the initial position of the center of the laser pulse.}
\label{sillage}
\end{figure*}

The first stage is simulated using the quasi-cylindrical PIC code CALDER-Circ \cite{lifs09}. We use a moving window of $3200\times200$ cells, with spatial steps $\Delta z = 0.25~c/\omega_0 $ and $\Delta r = 4~c/\omega_0$ longitudinally and transversely, and a time step $\Delta t = 0.249~\omega_0^{-1}$. We also use a pre-ionized plasma with 16 macro-particles per cell, and a scheme dedicated to the reduction of numerical Cherenkov radiation~\cite{lehe13}. A 15~J, 30~fs (FWHM) Gaussian laser pulse linearly polarized along the $x$ axis is focused on a 23~$\mu$m (FWHM) spot size at the entrance of the plasma, leading to a normalized peak vector potential $a_0=6$. The laser wavelength is $\lambda_0 = 800$~nm and the plasma has a density $n_e = 1.75\times 10^{18}$~cm$^{-3}$ with a linear entrance ramp of $200~\mu$m. This density is chosen so as to maximize the electron acceleration given the scaling laws of the blowout regime~\cite{lu06, lu07}, which is achieved when the depletion length equals the dephasing length $L_\mathrm{deph} = 15.3$~mm. A maximum energy gain of $\Delta E_{\textrm{max}} = 1.96$~GeV can then be expected. With these parameters, the high-intensity laser fields generate a strong and stable ion cavity [Fig. \ref{sillage}(a)] in which electrons can be accelerated to high energies. After about 15~mm of propagation, the drop of the normalized peak vector potential $a_0$ of the pulse indicates that the laser energy is depleted [Fig. \ref{sillage}(b)], and the laser can no longer drive the wakefield. The simulation shows that a mono-energetic component peaked at about 1.8~GeV is reached after 15~mm of propagation, in good agreement with the expected theoretical values. The corresponding injected charge is 5~nC above 350~MeV. After this distance, the electron beam starts losing energy by creating its own wakefield, as there is a natural transition to a beam-driven regime~\cite{pae10, cord11, mass14}. We then extract the electron bunch from the CALDER-Circ simulation when its energy is maximal after 15~mm of propagation.

In order to include non-symmetrical effects such as hosing instability, the second stage is simulated with CALDER in a Cartesian 3D geometry \cite{lefe03}, which is possible with manageable numerical cost. The low energy part ($<350$~MeV) of the beam injected in the simulation box is cut, which has negligible impact on the beam propagation and on the emitted radiation. The simulation box is now $800\times200\times200$ cells, with one particle per cell, a longitudinal step $\Delta z = 0.5~c/\omega_0 $, transverse steps $\Delta x = \Delta y = 0.5~c/\omega_0$ and a time step $\Delta t = 0.288~\omega_0^{-1}$. In this simulation, the plasma density is $n_e = 1.1\times 10^{20}~$cm$^{-3}$, about two orders of magnitude higher than in the first stage in the LWFA regime, with a very short entrance ramp of $25~\mu$m. Note that in the transition to the second stage simulation, the remaining laser fields are not registered and are thus completely suppressed. This physical approximation can be made because the remaining laser can not drive a wakefield on a significant distance in the second stage. This is justified for two reasons: (i) the laser pulse is strongly weakened at the end of the LWFA stage [see Fig. \ref{sillage}(b)] with $a_0\sim 3$ and 85\% of its energy already depleted after 15~mm, and (ii) the laser depletion length $L_\mathrm{d} \propto 1/n_e$ is very short in the high density plasma of the PWFA stage ($L_\mathrm{d}\sim100~\mu$m, while the electron bunch propagates over a few millimeters in this stage). In order to generate a wakefield in the blowout regime, the electron beam needs to have a density $n_{\mathrm{beam}}>1.8n_e$~\cite{lu06}. With our parameters, this lead to $n_e \lesssim 2\times 10^{20}~$cm$^{-3}$, limiting the density of this stage and leading to the chosen value. Note that the transverse size of the electron beam needs to remain below $\lambda_p$, which might also limit the plasma density used in the second stage. Due to its strong current (tens of kA), the accelerated electron beam generates its own wakefield [Fig. \ref{sillage}(c)] when propagating in the high-density plasma over about 3~mm. Contrarily to the previous blowout regime, the shorter plasma wavelength ($\lambda_p \sim 3$~$\mu$m at this high density) implies that the driving electron beam ($\sim 30~\mu$m long) overlaps with several ion cavities. As a consequence, the electrons situated in the front of each cavity are in a decelerating zone and lose energy by creating the wakefield, whereas the electrons at the rear of the cavities gain energy from the wakefield. This leads to the modulation of the longitudinal phase space of the electron bunch observed in Figure \ref{sillage}(d). The maximal energy of the electron bunch leaps up from 2~GeV to about 4~GeV in 740~$\mu$m of propagation in the plasma, indicating high-amplitude accelerating fields above 2500~GeV/m, which shows the potential of this regime in terms of boosting the electron energies to enhance the radiation emission. Eventually once the driving electrons at the front of the cavity have lost most of their energy, the wakefield slips backwards and the previously accelerated electrons take over in driving the wakefield. This lengthens the distance during which the electron beam will be able to generate a wakefield and perform betatron oscillations, which further optimizes the radiation emission.

\begin{figure}[]
\centering
\includegraphics[width=0.45\textwidth]{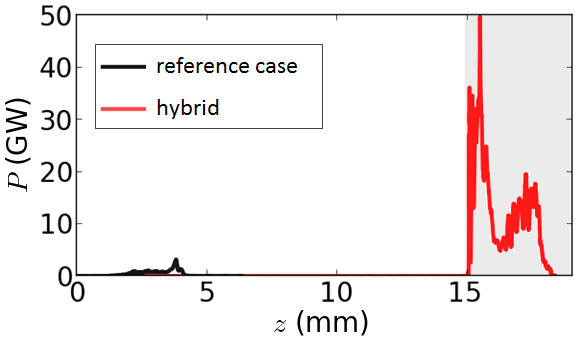}
\caption{Increase of the radiated power of the betatron source in the second stage. Instantaneous radiated power $P$ in the reference case (single stage at $n_e= 1\times 10^{19}$~cm$^{-3}$, solid black line) and in the two-stage hybrid case (solid red line). The light gray zone after 15~mm corresponds to the PWFA stage for the hybrid scheme.}
\label{pray}
\end{figure}

In Figure \ref{pray}, we show the effect of the high density of the second stage on the instantaneous betatron radiated power $P$. Results in our two-stage scheme are compared with a 3D CALDER simulation of a single-stage LWFA scheme using the same laser pulse, but at a constant plasma density $n_e = 1\times 10^{19}$~cm$^{-3}$, in the following referred to as the reference case. This intermediate density is chosen to directly optimize the betatron emission in a single stage. As the density is higher than in the first stage of the two-stage scheme, the shorter dephasing length leads to a less energetic electron bunch (quasi-monoenergetic bunch at 800~MeV). Moreover, the density is also lower than in the second stage (divided by 11), so electron wiggling is reduced. The betatron radiated power stagnates below a few GW between 2 and 4~mm, when the electron energy is close to its maximum (Fig. \ref{pray}). No radiation is then observed due to laser and beam depletion. In the two-stage scheme, $P$ is very low during the first stage, where it remains below 30~MW. However it then strongly increases by about 3 orders of magnitude in the second stage due to the density step. The radiated power reaches a maximal value of about 50~GW (Fig. \ref{pray}), thus exceeding the reference case by one order of magnitude. This can be justified considering that $P\propto \gamma^2n_e^2r_{\beta}^2$. If the characteristic length $L_\mathrm{ramp}$ of the plasma density entrance ramp of the second stage is small ($L_\mathrm{ramp}<\lambda_{\beta}\sim 300~\mu$m for 2 GeV electrons in the second stage, with $\lambda_{\beta} = \sqrt{2\gamma}\lambda_p$), $r_{\beta}$ won't be significantly changed, and the ratio of the radiated power between the two stages will be $P_2/P_1 = (n_2/n_1)^2$. In contrast, in the adiabatic case ($L_\mathrm{ramp}>\lambda_{\beta}$), the electron beam size is reduced in the entrance ramp and the ratio of the radiated power between the two stages reads $P_2/P_1 = (n_2/n_1)^{3/2}$.

\begin{figure}[]
\centering
\includegraphics[width=0.5\textwidth]{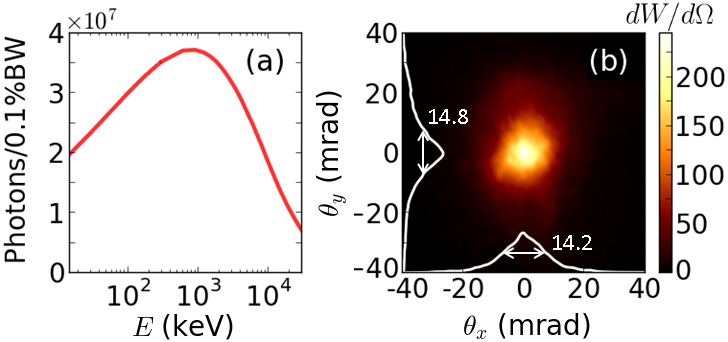}
\caption{Generation of a MeV betatron source in the two-stage hybrid scheme. (a) Photon spectrum after 3.3~mm of propagation in the PWFA stage (emission over a $60\times60$~mrad$^2$ solid angle centered on axis). (b) Angular energy distribution $dW/d\Omega$ (J/sr) of this source. On-axis lineouts are added in white with the FWHM value of the divergence.}
\label{Xray}
\end{figure}

By combining a laser-driven plasma accelerator and a beam-driven plasma radiator, we can then separate the acceleration process and the radiation emission process, which considerably boosts the betatron source. This is different from earlier ideas, which proposed a descending density step to improve the maximal energy of the electron beam~\cite{hidd10}. Besides, the generation of a plasma wake by an electron beam issued from a laser-plasma accelerator has been experimentally demonstrated recently~\cite{chou16}. Other methods based on plasma manipulations and trying to achieve a radiation enhancement by an increase of the oscillation radius have been proposed~\cite{taph08}, but were never implemented successfully. Figure \ref{Xray}(a) shows the photon spectrum of our hybrid betatron source after 3.3~mm of propagation in the PWFA stage. At this point, 90~\% of the electron bunch energy has been transmitted to the background plasma through wakefield excitation. The betatron emission peaks at a photon energy of about 1~MeV, and by fitting the spectrum with a synchrotron distribution given by the function $S(x) = x\int_x^{\infty}K_{5/3}(\xi)d\xi$ -- with $K$ a modified Bessel function of the second kind -- we can determine a critical photon energy $E_c = 9$~MeV. This is a much higher critical photon energy than the 245~keV value obtained in the reference case (with a peak at 30~keV). Besides, the total energy contained in the photon beam reaches 140~mJ, against $7.5$~mJ in the reference case. This in turn yields an energy conversion efficiency from the laser energy to the radiated energy as high as 0.9~\%. We also show in Figure \ref{Xray}(b) the angular distribution of the emitted $\gamma$-rays. We find a $14\times 15$~mrad$^2$ FWHM divergence, and assuming a $30~\mu$m-long and $2~\mu$m-wide electron bunch (FWHM), this leads to the estimated brilliance $ B = 4.4\times 10^{23}$~photons/s/mm$^2$/mrad$^2$/0.1\%BW at 1~MeV.
This high brilliance at an unprecedented high photon energy constitutes a considerable improvement compared with previous tabletop betatron or Compton sources from LWFA~\cite{knei10, chen13, sarr14}, with a potential increase of the source efficiency by several orders of magnitude.

In conclusion, the proposed scheme enables optimization of the betatron source at constant laser energy through the possibility of independent control of the electron acceleration and the strength of the electron wiggling, taking advantage of the PWFA regime. Thus, a strong increase by a factor of almost 40 of the critical photon energy of the radiation is observed, together with a significant improvement of the source efficiency. Emission in the beam-driven regime indeed enables such an increase of the energy conversion efficiency, as no additional source of energy is needed. The only restriction is that the electron beam must be able to drive a plasma wakefield in the blowout regime in the high-density stage. This requires high current and small transverse size to reach a beam density higher than the plasma density of the second stage. It has been shown to be verified in our study with an ideal 0.5~PW laser, and should soon be fully achievable in future multi-petawatt installations~\cite{papa13,eli,sung16,gan17}.
Finally, the improvement observed for the critical energy should enable the emission in the $\gamma$-ray domain with sub-PW lasers, which is very promising for numerous applications, as diverse as probing dense matter through gammagraphy~\cite{glin05} or detection of isotopes for homeland security~\cite{albe16}.

This work has been supported by the European Research Council (ERC) under the European Union's Horizon 2020 research and innovation programme (M-PAC project, grant agreement No. 715807) and by Laserlab-Europe (EU-H2020 654148). We also acknowledge GENCI for awarding us access to TGCC/Curie under the grant No. 2016-057594.

\end{document}